\documentclass[aps,prl,reprint,groupedaddress]{revtex4-1}

\usepackage{graphicx}
\usepackage{dcolumn}
\usepackage{bm}

\begin{document}


\title{Chern pump: a bridge between integer quantum Hall effect and quantum spin Hall effect}


\author{Yi-Dong Wu}
\email{wuyidong@ysu.edu.cn}
\affiliation{Department of Applied Physics, School of Science, Yanshan University, Qinhuangdao, Hebei, 066004, China}


\date{\today}

\pacs{73.43.-f, 72.25.Hg, 75.10.Pq, 85.75.-d}


\begin{abstract}
We propose a electron-pumping mechanism called Chern pump to explain the integer quantum Hall effect(IQHE) in the Chern insulator. By using the parallel transport gauge in the hybrid Wannier representation we establish the bulk and edge states correspondence in the Chern insulator. The same correspondence can also be established in two dimensional(2D) topological insulator(TI). So we can consider 2D TI as two time reversal(TR) related Chern insulators put together. The quantum spin Hall effect(QSHE) can be viewed as two TR related Chern pumps pumping electrons to opposite directions. Compared with the $Z_2$ spin pump, the two Chern pumps explanation of QSHE is inherently 2D and predict that the QSHE can be detected in isolated device, thus make the QSHE directly measurable.
\end{abstract}

\maketitle
The quantization of the Hall conductance in integer quantum Hall effect has been explained from different perspectives since its discovery. Laughlin's gauge argument played a major role in the development of the theory of IQHE \cite{Laughlin1981}. He considered the two dimensional electron gas as a electron pump. Integer number of electrons are pumped from one edge of the cylinder to another when the magnetic flux threading the cylinder varies by one flux quantum $hc/e$. Halperin refined Laughlin's argument and stressed the role of the nonlocal edge states\cite{Halperin1982}.\\
 Thouless, Kohmoto, Nightingale and den Nijs calculated the Hall conductance of 2D electron gas in uniform magnetic field and period potential by using Kubo formula\cite{Thouless1982}. The precise quantization of the Hall conductance of the bulk 2D system is related to the topological invariant called TKNN number or Chern number of the occupied energy bands on the magnetic Brillouin zone. Haldence proposed an insulator with occupied band that has nonzero Chern number even in the absence of macroscopic magnetic field\cite{Haldane1988}. We refer those materials as ``Chern insulators" (CI). In this letter we show CI can also be considered as electron pump.\\
Recently quantum spin Hall effect is proposed in 2D topological insulators(TIs)\cite{Kane2005a,Kane2005b,Bernevig2006,Konig2007,Roth2009,Fu2006}. The 2D TI has gapless edge states on the boundary while the bulk is insulating. Kane and Mele proposed a $Z_2$ topological invariant to characterize the gapless edge states\cite{Kane2005b} and a $Z_2$ adiabatic spin pump to explain the QSHE in 2D TI\cite{Fu2006}. Though the $Z_2$ pump picture of QSHE has been widely accept by the community, we find that it may not be the true story. The claim ``an isolated $Z_2$ pump returns to its original state after two cycles" is especially doubtful, which means QSHE can never been observed in isolated device. In this letter we explore the similarity of the QHE in CI and QSHE in 2D TI and explain why the QSHE can be realized in isolated 2D TIs.\\
First we show that CI can be considered as an electron pump. We consider a cylinderical configuration as indicated in Fig. 1(a). $\mathbf{a_1}$ and $\mathbf{a_2}$ are the primary vectors of the 2D lattices, and $\mathbf{a_2}$ is parallel to the edges of the cylinder. $\mathbf{b_1}$ and $\mathbf{b_2}$ are the primary vectors of the reciprocal lattices which satisfy the $\mathbf{a_i}\cdot \mathbf{b_j}=2\pi\delta_{ij}$. We use the Berry-phase definition of the bulk polarization of CI as proposed in ref \cite{Coh2009}
\begin{equation}
\bar{\mathbf{r}}_n=i\frac{S}{(2\pi)^2}\int_{[\mathbf{k_0}]}d\mathbf{k}\langle u_{n\mathbf{k}}|\nabla_{\mathbf{k}}|u_{n\mathbf{k}}\rangle
 \end{equation}
$|u_{n\mathbf{k}}\rangle$ are the cell-periodic Bloch functions, $S$ is the cell area and $[\mathbf{k_0}]$  indicates the parallelogram reciprocal-space unit cell with origin at $\mathbf{k_0}$. This definition of the bulk polarization depend on the choice of $\mathbf{k_0}$. To see this, we focus on the polarization perpendicular to the edges.
\begin{equation}
\overline{x}_n=\frac{1}{2\pi}\int_{k_{20}}^{k_{20}+2\pi}dk_2 l\Phi_n(k_2)
 \end{equation}
\begin{equation}
\Phi_n(k_2)=i\frac{1}{2\pi}\int_{k_{10}}^{k_{10}+2\pi}dk_1\langle u_{n\mathbf{k}}|\frac{\partial}{\partial k_1}|u_{n\mathbf{k}}\rangle
 \end{equation}
where $k_i=\mathbf{k}\cdot\mathbf{a_i}$, $k_{i0}=\mathbf{k_0}\cdot\mathbf{a_{i}}$ and $l$ is the distance between two adjacent lattice lines as indicated in Fig. 1(a). $\Phi_n(k_2)$ gives the x position(in unit of $l$) of the one dimensional(1D) Wannier function center for a given $k_2$. Here, we have used the hybrid Wannier representation as in ref \cite{Soluyanov2011}. $\Phi_n(k_2)$, defined modulo $1$, is gauge independent and the continuous variation of $\Phi_n$ when $k_2$ increases by $2\pi$ defines the Chern number $C_n$ of the band\cite{Bohm2003}. we can see $\overline{x}_n$ is just the average x position of the hybrid Wannier function centers. When the $C_n$ is nonzero, $\overline{x}_n$ depends on $k_{10}$. However, as pointed out in ref\cite{Coh2009}, the adiabatic change of the polarization is independent of $\mathbf{k_0}$. For example, when the magnetic flux threading the cylinder increase by one flux quantum $\phi_0=hc/e$ a 1D hybrid Wannier state with $k_2$ will evolve into a state with $k_2+2\pi/N$, so $\overline{x}_n$ will change by $C_nl/N$($N$ is the number of the lattices along $\mathbf{a_2}$ direction). In this process, the Hamiltonian experiences a cycle and $C_n$ electrons pass through a cross section of the cylinder, that is, $C_n$ electrons are pumped, we refer to this process as a Chern pump.\\
Intuitively, the existence of the gapless edge states is inevitable. There must be room to accommodate the pumped electrons and the holes left at the other edge. If the edge states are gaped and the Fermi level is in the gap, pumping electrons from one edge to another will cost a large mount of energy in an isolated device, which can't be achieved by adiabatic change of Hamiltonian in one cycle.\\
 The relationship between edge states and bulk property in quantum Hall systems and TIs has been discussed in several works\cite{Halperin1982,Kane2005b,Fu2006,Hatsugai1993}. Halperlin calculated the energy of electron gas near the edge by using vanishing boundary conditions at the edges\cite{Halperin1982}. The radial wave functions of edge states and the bulk states have same number of nodes. When the magnetic flux vary the edge and bulk states can continuously evolve to each other. We can take the view that the bulk states are pushed upward by the edge potential to form the gapless edge states.\\
Recently, in a brilliant work the topology of the edge states in insulator is identified with that of the spectrum of a certain gluing function by using a simple model for the boundary between the insulator and vacuum\cite{Fidkowski2011}. In the 2D case for a given $k_2$ the gluing functions are just the centers of the 1D general Wannier functions when using the parallel transport gauge\cite{Vanderbilt1993,Kivelson1982}. Those 1D general Wannier functions span the same space as the occupied Bloch bands and are maximally localized in 1D real space. \\
\begin{figure}
  \includegraphics[width=3.7 in,clip=true]{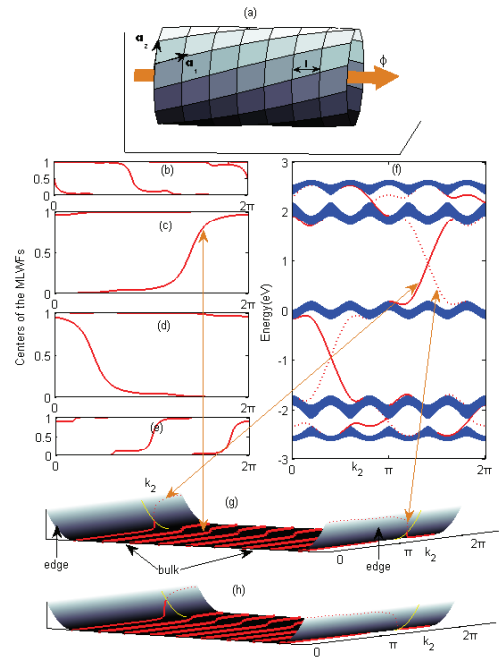}\\
  \caption{(a)The geometrical configuration of the 2D insulator. (b)-(e) The centers of 1D MLWFs $\Phi_n$s as functions of $k_2$ when four,three,two and one band(s) is(are) occupied with $q/p=2/5$ in the tight-binding model. The $\Phi_n$s of the trivial bands are constantly zeros module one, so they aren't shown. (f)The energy spectrum  of the tight-binding model. the red lines are the edge states, solid lines on one edge and the doted lines on the other. The correspondence between the $\Phi_n$s and the edge states is obviously. (g)An intuitive illustration of how the 1D maximally localized Wannier states are pushed upward to form the gapless edge states. The solid lines at edge are occupied and the doted lines are unoccupied. The magnetic flux $\phi=0$ ,Fermi levels of the two edge are equal. The system is in its ground state. (h)An illustration of  the change of occupation of the edge states when the magnetic flux varies from zero to a finite value. With the electrons pumped from one edge to the other, the Fermi levels of the two edges become different. }\label{fin}
\end{figure}
Using the centers of the maximally localized Wannier functions(MLWF) $\Phi_n$ as function of $k_2$ to characterize the edge state has several advantages. First, it's a very general conclusion, which can be applied in  both TR symmetric and broken insulators. Second, it's more accurate in characterizing edge states than Chern numbers or the $Z_2$ invariants. The actual boundary conditions of insulators are quite complicated, we don't expect such a simple model can accurately determine the shape of the edge states. However, as indicated in Fig 1., there is a remarkable geometrical resemblance between the $\Phi_n$s and the edge states calculated with the tight-binding model in ref\cite{Hatsugai1993}. We have calculated the centers of the 1D MLWFs of 2D, 3D TIs and topological semi-metals\cite{Wu2011a}. We find they accurately predict the topological properties and even provide some geometrical information of the edge states. Finally, it provides an intuitive view of how the edge states are formed at the boundaries of the material as illustrated in Fig.1.(g). Similar to the discussion of electron gas in ref\cite{Halperin1982}, in the CI or TIs we can think it's the maximally localized hybrid Wannier states that are pushed upward to form the gapless edge states as illustrated in Fig. 1.(g). In this way we establish the bulk and edge states correspondence. When magnetic flux varies they can adiabatically evolve to each other .\\
 In Fig. 1(h) we demonstrate the adiabatic evolution of the hybrid Wannier states and the edge states when the magnetic flux varies. The centers of the MLWFs can be smoothly jointed at $k_2=0$ and $k_2=2\pi$, so the hybrid Wannier states in bulk remain occupied. However, the filling of the edge states will change adiabatically and there will be a difference in Fermi levels of the two edges( because of the net charges at the edges, there will also be electrostatic potential between the two edges). Thus, we can create a quantum Hall state by adiabatic evolution when the device is isolated. By ``isolated" we mean there are no reservoirs connected to the edges of the cylinder.\\

Once how the CIs pump electrons is understood, the mechanism of the QSHE in 2D TIs becomes clear. The only difference between TIs and CIs is the TIs respect TR symmetry. 
 Because of the TR symmetry, the hybrid Wannier functions constructed by sing the parallel gauge in TR symmetric insulators are also TR related, that is, the MLWFs at $k_2$ and $-k_2$ are related by TR operator\cite{Fidkowski2011}. Transformed back to the Bloch representation, the two group of 2D bands are also TR related. Because the $\Phi_n$s are equal at $k_2$ and $-k_2$ as indicated in Fig. 2, the two group of bands have opposite Chern numbers $\pm C$. $C$ is odd in 2D TI and even in trivial insulator. In fact, the occupied bands of 2D TIs can only be decomposed to two groups of bands with odd Chern number if we insist the two group of bands are TR related\cite{Wu2011b}. In contrast, the trivial insulator can be decomposed to two groups TR related trivial bands with zero Chern number. So $Z_2$ is just the parity of the Chern number of one of groups of bands decomposed using the parallel transport gauge.\\
  The bulk and edge states correspondence discussed in the Chern insulators is still valid here, two groups TR related maximally localized hybrid Wannier states correspond to two branches of TR related edge states as indicated in Fig. 2. So 2D TI can be viewed as two TR related Chern insulators put together. If we consider a geometrical configuration in Fig.1.(a) with 2D TI, when the magnetic flux change one flux quantum, the two TR related CIs will each pump $C$ electrons to the opposite directions. If the 2D TI is initially in its ground state as indicated in Fig 2, there will be $C$ electrons added and $C$ holes left at each of the two edges after one pump. If the average values of spins of the edge states are nonzero, net spin will accumulate at the edges after one pump. In this way, the spins are pumped from one edge to another. However, because spins aren't conserved in most of the materials, so in general we can't get an integer quantum spin Hall conductance as in ref\cite{Bernevig2006}. When the magnetic flux changes continuously the two Chern pumps will continuously pump electrons to opposite directions. If the device is isolated the difference of the Fermi levels of the two branches edge states at each edge will increase as indicated in Fig 2.(f).\\
\begin{figure}
  \includegraphics[width=3.7 in,clip=true]{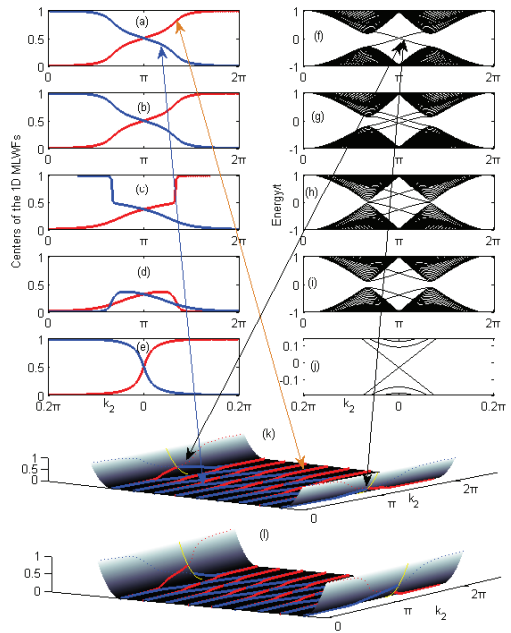}\\
  \caption{(a)-(d)The centers of MLWFs $\Phi_n$s with the tight-binding model in ref\cite{Kane2005b}. The parameters are $\lambda_{SO}=0.06t$, $\lambda_R=0.05t$ and $\lambda_v=0,0.1t,0.29t,0.4t$ respectively. They illustrate a TI and normal insulator transition. (f)-(i) show the corresponding energy spectrum of the model. The lines at the bulk gap are edge states. They have topology of the $\Phi_n$s. (d) and (j) show the results of HgTe quantum well by using the tight-binding model in ref\cite{Bernevig2006}. To make the edge state more localized we use a parameter $M=-0.1$. (k) An illustration of the formation of edge states similar to the CI case in Fig.1.(1) The occupation of the edge states when electrons are pumped by the two Chern pumps to opposite directions. Net spins and currents present at the two edges. }\label{s2d}
\end{figure}
Now we discuss why we prefer the two Chern pumps to the $Z_2$ adiabatic spin pump proposed in ref\cite{Fu2006} in explaining the QSHE. Our first objection to use the $Z_2$ pump is that the $Z_2$ pump isn't a primary pump in the 2D system. By definition a pump means that the Hamiltonian of the system experiences a cycle and returns to its origin form. However, because $Z_2$ pump is defined in a 1D system, the external parameter $t$ or $k_2$ must varies by $2\pi$ to finish a cycle. In a 2D system illustrated in Fig 1.(a), this means the Hamiltonian returns to its origin form $N$ times. In a general 2D system, e.g. a Corbino disc, when the $k_2$ can't be defined, it isn't clear how many times the Hamiltonian returns to its origin form to achieve a $Z_2$ pump. Even in the special case that the $Z_2$ pump can be defined, the claim that ``an isolated $Z_2$ pump returns to its original state after two cycles" can never be realized in real 2D system. In fig 2. after two $Z_2$ cycles $k_2$ will varies by $4\pi$, the electrons at the  edge state will adiabatically evolve to the bulk valence bands if the single particle picture remains valid.\\
 Contrary to the $Z_2$ pump picture, in our two Chern pumps picture we don't need coupling to the reservoir to explain why 2D TIs can pump spins. If coupling to the reservoirs must be include in the QSHE, the QSHE would become more or less an external property of the TI, which isn't aesthetically satisfactory. Besides, the interactions between spins and reservoirs are very complicated in general, which make it difficult to observe the QSHE experimentally. In our opinion, the edge states serve as natural reservoirs to accommodate the spins pumped by bulk energy bands in an isolated device. \\
\begin{figure}
  \includegraphics[width=2 in,angle=180,clip=true]{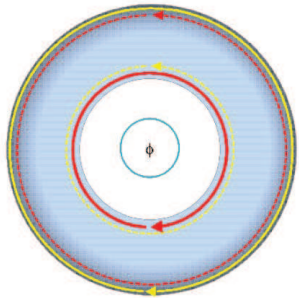}\\
  \caption{A suggested device to test the TR Chern pump picture. The Corbino disc is constructed from HgTe quantum well. The device is isolate. By varying the magnetic flux $\phi$ spins are pumped to the edges. The red(yellow) lines are the spin up(down) edge state. Solid lines have higher Fermi level than the dashed lines. The arrows denote the direction of the speed of the electrons. There are net up(down) spins at the inner(outer) edge. The directions of the currents are same at the two edges(opposite to the higher Fermi level electrons).}\label{disc}
\end{figure}
The most outstanding advantage of two TR Chern pumps picture is that it make the quantum spin Hall effect directly measurable. So far the experimental researches on the QSHE have focused on the existence of the nonlocal edge states\cite{Konig2007,Roth2009,Fu2006}. Only recently are the edge states confirmed to be spin-polarized\cite{Brune2012}. However, no QSHE has been directly observed, that is, no net spin accumulation at the edge of the 2D TI caused by the traverse electric field has been detected.\\
 Since we view the 2D TI as two TR related CIs and the Hall conductances of the CIs aren't affected by the geometrical configuration of the device, the 2D TI can still be considered as two Chern pumps in a Corbino disc as indicated in Fig.3. The bulk band and the edge states correspondence guarantees the existence of the TR related gapless edge states. So when the magnetic flux varies the two Chern pumps will pump electrons to opposite directions, spins will accumulate at the edges. In this process the spins are transported through bulk instead of the edge channels, so we don't have to worry about the complicated interaction between the spin currents at the edges and the reservoirs. \\
  We can use HgTe quantum well to construct such a device and spins accumulated at the edge will be out of plane. This is a very interesting state, the Fermi levels of spin up and down electrons are different at each edge. So there are electric currents at the edges and the currents are in the same direction. This state can be detected either by measuring the spins at the edge directly or by measuring the edge current.\\
   For a constant magnetic flux, if the single particle picture remains valid and there are no TR symmetry breaking impurities, the current at the edge will be nondissipative. In this sense, the edge states are ideal reservoirs for the pumped spins since the spin-polarized states at the edges have long if not infinite relaxation time. However, when these states are far from equilibrium, the many-particle effect will certainly cause some relaxation. Therefore, with our proposition not only can the pumped spins be detected but also the relaxation of the pumped spins can be studied experimentally.\\
In conclusion, we proposed a spin pump in 2D TI in analogy to the Laughlin's charge pump or the Chern pump in CI. We consider the 2D TI as two TR related CIs which can pump electrons to opposite directions and the edge states serve as natural and ideal reservoirs for the pumped spin.  In this way, we establish a coherent physical picture of the QSHE and make the QSHE directly measurable.



%

\end{document}